\documentclass[journal]{IEEEtran}

\usepackage[numbers,sort&compress]{natbib}
\usepackage{amsmath}
\usepackage{amssymb}
\usepackage{graphicx}
\usepackage{epstopdf}

\usepackage {graphicx}      %picture
\graphicspath{{fig/}}
\DeclareGraphicsExtensions{.eps}
\usepackage{epstopdf}

\begin{document}
\title{Low-Complexity Sphere Decoding of Polar Codes based on Optimum Path Metric}
\author{Kai~Niu,~Kai~Chen~and~Jiaru~Lin
\thanks{This work was supported by National Natural Science Foundation of China (No. 61171099) and Qualcomm Corporation.}
\thanks{The authors are with the Key Laboratory of Universal Wireless Communication, Ministry of Education,
Beijing University of Posts and Telecommunications,
Beijing 100876, China. (e-mail: {niukai, kaichen,jrlin}@bupt.edu.cn)}% <-this % stops a space
%\thanks{Digital Object Identifier }
}

%\markboth{IEEE COMMUNICATIONS LETTERS,~Vol.~X, No.~X, May~2012}%
%{Shell \MakeLowercase{\textit{et al.}}: Bare Demo of IEEEtran.cls for Journals}

\maketitle

\newtheorem{theorem}{Theorem}
\newtheorem{example}{Example}
\newtheorem{algorithm}{Algorithm}
\begin{abstract}
%\boldmath
Sphere decoding (SD) of polar codes is an efficient method to achieve the error performance of maximum likelihood (ML) decoding. But the complexity of the conventional sphere decoder is still high, where the candidates in a target sphere are enumerated and the radius is decreased gradually until no available candidate is in the sphere. In order to reduce the complexity of SD, a stack SD (SSD) algorithm with an efficient enumeration is proposed in this paper. Based on a novel path metric, SSD can effectively narrow the search range when enumerating the candidates within a sphere. The proposed metric follows an exact ML rule and takes the full usage of the whole received sequence. Furthermore, another very simple metric is provided as an approximation of the ML metric in the high signal-to-noise ratio regime. For short polar codes, simulation results over the additive white Gaussian noise channels show that the complexity of SSD based on the proposed metrics is up to $100$ times lower than that of the conventional SD.
\end{abstract}


\section{Introduction}

\IEEEPARstart{A}s a major breakthrough in coding theory, polar codes, introduced by Ar{\i}kan in \cite{pcode}, can asymptotically achieve the channel capacity of binary symmetric channels using a successive cancellation (SC) decoder with a complexity of $O\left( N\log N \right)$, where $N={{2}^{n}},n=1,2,...$, is the code block length.
Later, some improved algorithms of SC are described in \cite{scl} \cite{scl2} \cite{scs}.
Yet still the performance of these decoders is inferior to that of the maximum likelihood (ML) decoder, or at least cannot be proven to achieve the ML performance.
An ML decoder of polar codes is implemented by Viterbi algorithm in \cite{ml}; but the decoding complexity grows exponentially with the code length.

Inspired by the sphere decoding (SD) for space-time block codes \cite{lcdec} \cite{mlclp}, an SD algorithm of polar codes is proposed in \cite{psd} to perform ML decoding with a cubic complexity. The sphere decoder configures an initial target radius, and the candidates in the sphere are enumerated.
Whenever an $N$-length path corresponding to a smaller radius is found, this path is buffered as a candidate codeword, and based on which, the target radius is updated.
After that, the search is restarted in the new sphere with a smaller radius. Until no paths can be found in the sphere with a smaller radius than the target, the decoder outputs the buffered path as the decoding codeword.
The drawback of this naive decoding is that some paths which have already been spanned in the previous sphere may be respanned.
Furthermore, when searching in the sphere, the decoder enumerates and expands the candidate path only based on the path length, namely, if one path has a longer length, it will be expanded with a higher priority. This length-first search is not a good strategy, because the search rule only involves a simple information of the path length and neglects the information provided by the received sequence.

Stack sequential algorithm, as stated in \cite{mlclp}, is functionally equivalent to the sphere decoding. In this paper, a stack SD (SSD) algorithm of polar codes is applied, where the paths are stored in a stack with a descending order of the path metrics. In SSD, each possible path is visited at most once. Inspired by Massey's work about the optimum frame synchronization \cite{ofs}, we derive an optimum path metric based on the ML rule. Further, a much simpler metric is also provided as an approximation of the proposed metric in the high signal-to-noise ratio (SNR) regimes. With the help of the proposed path metrics, the SSD can efficiently enumerate the candidate paths and narrow the search range so as to remarkably reduce the computational complexity yet still achieve the ML performance.

The remainder of this paper is organized as follows. \mbox{Section \ref{secII}} describes the process of the polar coding and the signal model over the additive white Gaussian noise (AWGN) channel. \mbox{Section \ref{secIII}} derives the optimum and suboptimum path metric based on the ML rule and presents the details of SSD algorithm. \mbox{Section \ref{secIV}} gives the performance and complexity analysis of the proposed SSD algorithm based on the simulation results. Finally, \mbox{Section \ref{secV}} concludes the paper.

\section{Polar coding and signal model}
\label{secII}
Given an $\left( N,K \right)$ polar code, the $N$-length code block (codeword) $\mathbf{x}$ can be generated by
\begin{equation}
    \mathbf{x}=\mathbf{v}{{\mathbf{F}}}
\end{equation}
where $\mathbf{v}={{\left( {{v}_{1}},{{v}_{2}},...,{{v}_{N}} \right)}}=\mathbf{u}\mathbf{B}$ is a source-scrambling vector generated by the $N \times N$ bit-reversal permutation matrix ${{\mathbf{B}}}$ defined in \cite{pcode}, and the vector $\mathbf{u}\in {{\left\{ 0,1 \right\}}^{N}}$ is the source block.
As the $n$-th Kronecker power of $\left[ \begin{matrix}
   1 & 0  \\
   1 & 1  \\
\end{matrix} \right]$, the matrix ${{\mathbf{F}}}$ has a lower triangular structure, where the $(j,i)$-th element $f_{ji}$ is taken the value in $\{0,1\}$.
Hence, the components $x_i\in \mathbf{x}$ can be written as
\begin{equation}
x_i  = \sum\limits_{j = i}^N {f_{ji} v_j }.
\end{equation}
The $N$ polarized subchannels are corresponding to the rows of the matrix ${{\mathbf{F}}}$, and the reliabilities of which are calculated using Bhattacharyya parameters \cite{pcode}.
The entries in the source-scrambling vector $\mathbf{v}$ with indices corresponding to the more reliable $K$ rows of the matrix $\mathbf{F}$ are assigned information bits; the other $N-K$ entries are related to the frozen bits which can be set to zeros when the channels are symmetric.

The received sequence $\mathbf{y}={\left( {{y}_{1}},{{y}_{2}},...,{{y}_{N}} \right)}$ can be written as
\begin{equation}
    \mathbf{y}=\sqrt{E}\mathbf{s}+\mathbf{w}
\end{equation}
where $E$ is the energy of the transmitted signal, and $\mathbf{s}={\left( {{s}_{1}},{{s}_{2}},...,{{s}_{N}} \right)}$ is the transmitted signal vector, and $\mathbf{w}$ is an $N$-length vector of i.i.d. AWGNs. The components $w_i\in\mathbf{w}$ are statistically independent Gaussian random variables with $0$ mean and variance ${N_0/2}$ where $N_0$ is the one-sided noise spectral density. And the elements $s_i=1-2x_i$ are the BPSK modulated signals.
The ML estimation of the transmitted bits can be obtained by minimizing the square Euclidean distance (SED) towards the received sequence, that is,
\begin{eqnarray}
    \hat{\mathbf{v}}&=&\mathop {\arg \min }\limits_{{\bf{s}} = {\bf{1}}_N  - 2{\bf{vF}}} \left| {{\bf{y}} - \sqrt{E}{\bf{s}}} \right|^2 \\
    &=& \underset{\mathbf{v}}{\mathop{\arg \min }}\,{{\left| \mathbf{y}-\sqrt{E}\left({{\mathbf{1}}_{N}}-2{\mathbf{v}{\mathbf{F}}}\right) \right|}^{2}}
\end{eqnarray} where $\mathbf{1}_N$ is an all-one vector of length $N$.
After performing a bit-reversal permutation on $\hat{\mathbf{v}}$, an estimation of the source block is obtained, i.e., $\hat{\mathbf{u}}=\hat{\mathbf{v}}\mathbf{B}^{-1}$ where $( .)^{-1}$ denotes the matrix inverse. In fact, $\mathbf{B}=\mathbf{B}^{-1}$.

\section{Stack sphere decoding and the path metrics}
\label{secIII}
In this section, we first give a brief description of the code tree and provide an SSD with conventional enumeration strategy.
Then, the optimum path metric is derived based on the ML rule and its approximations in high/low SNR regimes are also provided.
Finally, the proposed SSD algorithm with an efficient enumeration is described in detail.

\subsection{SSD with conventional enumeration}
For the SD algorithm, the ML estimation $\hat{\mathbf{v}}$ can be obtained by enumerating all the transmitted signal vectors $\mathbf{s}$ within a sphere of radius $r$ that is centered at $\mathbf{y}$.
In \cite{psd}, the SD algorithm is described as a series of depth-first searches on the code tree, and a strategy using parallel subtree searches is suggested to reduce the complexity.
But for each iteration of both methods, after updating the radius of sphere, the search process should be restarted from the preceding node.
Thus, for a direct implementation of SD, a certain path in the code tree may be revisited many times.
To avoid these revisitings, a stack implementation of SD is applied in this paper.

\begin{figure}[!t]
\centering{
\includegraphics[width=80mm]{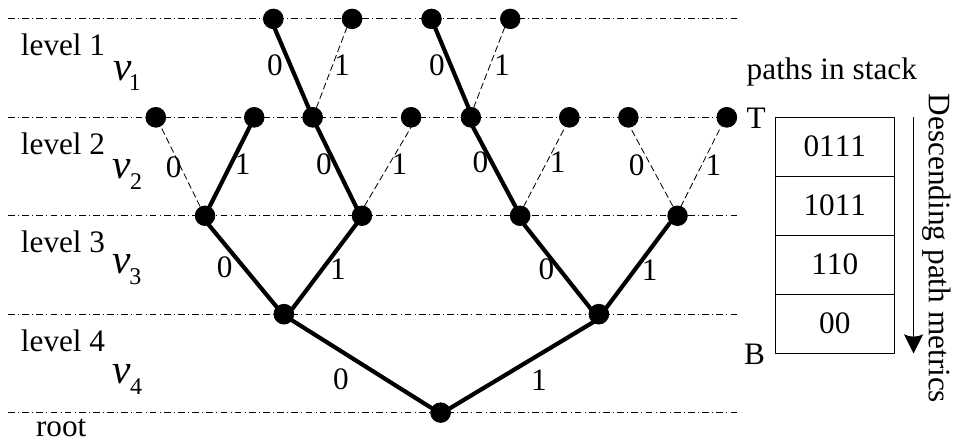}}
\caption{an example of code tree and paths in stack of SSD.}
\label{fig_example}
\end{figure}

For finite BPSK signals, the sphere decoder can be regarded as a bounded search in a code tree. Due to the lower triangular form of ${\mathbf{F}}$, the source-scrambling vector $\mathbf{v}$ can be represented as a path in a tree of depth $N$, where the values of bit $v_i$ at level $i$ correspond to branches at depth $N-i+1$. For example, Fig.\ref{fig_example} gives a code tree with four levels.
Inspired by \cite{scs}, a stack is used in SSD to store the candidate paths.
The paths with solid branches are the candidates in the stack and sorted by path metric in descending order, on the other hand, those dash branches are visited and discarded (since their Euclidean distances to the sphere center are larger than the target).
Under the SSD algorithm, a possible candidate path is visited at most once.

Let $v_i^N=\left( v_i,v_{i+1},...,v_N \right)$ denote the binary labeling of an $(N-i+1)$-length path from the root to a certain branch at the $i$-th level, and we use $s_i^N=\left( s_i,s_{i+1},...,s_N \right)$ to denote the corresponding transmitted signal vector.

Given the received sequence $\mathbf{y}$, the corresponding SED between the path $v_i^N$ and the related segment of received sequence $y_i^N$ can be expressed as that in \cite{psd},
\begin{eqnarray}
\label{equ_metric}
    D \left( {v_i^N } \right) &=& \sum\limits_{l = i}^N {\left| {y_l  - \sqrt{E}s_l } \right|^2 }  \\
\label{MSED_metric}
     &=& \sum\limits_{l = i}^N {\left| {y_l  -\sqrt{E}\left(1 - 2\sum\limits_{j = i}^N {f_{jl} v_j }\right) } \right|^2 }.
\end{eqnarray}
If the branch $v_i$ is corresponding to a frozen bit and takes an incorrect value, the associated SED is $D\left(v_i^N\right)=\infty$.
The sphere decoder utilizes this SED as a measurement to judge whether the vector $v_i^N\left (s_i^N\right)$ is inside the sphere. Particularly, when $i=1$, it is also used to update the radius of sphere.

For the conventional enumeration, since the SD is described as a series of depth-first searches, the length of the candidate path is taken for path sorting in SSD,
\begin{equation}
\label{M0_metric}
M_0\left( {v_i^N } \right) =N-i+1.
\end{equation}
Initially, a null path $\phi$ which is corresponding to the root node is loaded in the stack and its metric is $M_{0}\left( \phi  \right)=0$.
The search goes along the longest path in the stack.
Whenever a candidate path $v_i^N$ is found to have a measurement $D\left(v_i^N\right)$ that is larger than the target radius $r$, it will be dropped directly.
When the path on the top of the stack achieves length $N$, i.e., $i=1$, the $N$-length bit sequence corresponding to this path is buffered as a candidate vector, and the target radius is updated as $r=\sqrt{D\left(v_i^N\right)}$.
This procedure goes on until the stack becomes empty, the buffered vector is then output as the decoding result.
In fact, this conventional enumeration is to select the candidate path in the stack one-by-one and check that whether the candidate is inside the sphere. Since the paths in the stack are sorted by the simple path metric $M_0$, the sphere decoder lacks the valuable information to direct the search procedure.

\subsection{Optimum path metric}
Given a path $v_i^N$ and its corresponding vector $s_i^N$, the complement vector $s_1^{i-1}$ can take a possible value from a random vector ${{d_1^{i-1} }} = \left( {d _1,d_2 ,...,d _{i - 1} } \right)$, where each $d_j,j=1,2,...,i-1$ is either $+1$ or $-1$ with an equal probability of $1/2$. The optimum path metric for the enumeration is to maximize the likelihood probability
\begin{equation}
\label{Metric_L1}
L_1  = \sum\limits_{\text{all }{d_1^{i-1}}} {P\left( {{\bf{y}}\left| {s_i^N s_1^{i - 1} } \right.} \right)P\left( {s_1^{i - 1}  = {d_1^{i-1}}} \right)}.
\end{equation}
Since $P\left( {s_1^{i - 1}  = {d_1^{i-1}}} \right) = 2^{ -\left( i-1\right)}$ for all ${d_i^{i-1}}$, it is equivalent to maximize
\begin{equation}\label{Metric_L2}
L_2  = \sum\limits_{\text{all }{d_1^{i-1}}} {P\left( {{\bf{y}}\left| {s_i^N } \right.} \right)P\left( {{\bf{y}}\left| {s_1^{i - 1}  = {d_1^{i-1}}} \right.} \right)}
\end{equation}
Since $\mathbf{w}$ is a vector of i.i.d. AWGNs, we have
\[
P\left( {{\bf{y}}\left| {s_i^N } \right.} \right)P\left( {{\bf{y}}\left| {s_1^{i - 1}  = {d_1^{i-1}}} \right.} \right) = \left( {2\pi } \right)^{ - N/2}\cdot\qquad\qquad\qquad
\]
\begin{equation}
 \prod\limits_{l = i}^N {\exp \left( { - \frac{{\left( {y_l  - \sqrt E s_l } \right)^2 }}{{N_0 }}} \right)} \prod\limits_{l = 1}^{i - 1} {\exp \left( { - \frac{{\left( {y_l  - \sqrt E d_l } \right)^2 }}{{N_0 }}} \right)}.
\end{equation}
Substituting this formula into (\ref{Metric_L2}) and neglecting all the items independent of $s_i^N$ or $d_1^{i-1}$, it is equivalent to maximize
\begin{equation}\label{Metric_L3}
L_3  = \sum\limits_{\text{all }{d_1^{i-1}}} {\prod\limits_{l = i}^N {\exp \left( {\frac{{2\sqrt E y_l s_l }}{{N_0 }}} \right)} \prod\limits_{l = 1}^{i - 1} {\exp \left( {\frac{{2\sqrt E y_l d_l }}{{N_0 }}} \right)} }.
\end{equation}
Remember $d_l$ takes value in $\{-1, +1\}$, the above summation can be expressed as
\begin{equation}
\label{Metric_L3e}
L_3  = \prod\limits_{l = i}^N {\exp \left( {\frac{{2\sqrt E y_l s_l }}{{N_0 }}} \right)} \prod\limits_{l = 1}^{i - 1} {\left(2\cosh \left( {\frac{{2\sqrt E y_l }}{{N_0 }}} \right)\right)}.
\end{equation}
The logarithmic form of (\ref{Metric_L3e}) is
\begin{equation}
L_4  = \sum\limits_{l = i}^N {\frac{{2\sqrt E y_l s_l }}{{N_0 }}}  + \sum\limits_{l = 1}^{i - 1} {\left(\log 2\cosh \left( {\frac{{2\sqrt E y_l }}{{N_0 }}} \right)\right)}.
\end{equation}
Given a received sequence $\mathbf{y}$, the following summation
\begin{equation}
\label{const}
\sum\limits_{l = 1}^N {\left(\log 2\cosh \left( {\frac{{2\sqrt E y_l }}{{N_0 }}} \right)\right)}
\end{equation}
is a constant. After subtracting (\ref{const}) from $L_4$, we obtain a path metric that is equivalent to the optimal one in (\ref{Metric_L1}), i.e.,
\begin{eqnarray}\label{M1_metric}
&& M_1 \left( {v_i^N } \right)=\sum\limits_{l = i}^N {\frac{{2\sqrt E y_l s_l }}{{N_0 }}}-h_1 \left( {y_i^N } \right)\nonumber\\
&&=\sum\limits_{l = i}^N {\frac{{2\sqrt E }}{{N_0 }}} y_l  \left(1 - 2\sum\limits_{j = i}^N {f_{jl} v_j }\right)-h_1 \left( {y_i^N } \right) \qquad
 \end{eqnarray}
where
\begin{equation}
\label{equ_h1}
h_1 \left( {y_i^N } \right)=\sum\limits_{l = i}^N {\log \cosh \left( {\frac{{2\sqrt E y_l }}{{N_0 }}} \right)} + \left( {N - i+1} \right)\log 2.
\end{equation}

This ML path metric includes two terms: the first summation involves the correlation between the vector $s_i^N$ and the received sequence $y_i^N$; the second term, $h_1\left( {v_i^N } \right)$, is a correction term which further consists of two parts: the summation represents a kind of energy correction related to the received sequence $y_i^N$; the second part is corresponding to the depth of the path $v_i^N$. Compared with Massey's metric \cite{ofs}, this metric has the same item for the first summation, but the correction term of which includes an additional part to manifest the length of the path.

Similar to the works in \cite{ofs}, two approximations of the optimum path metric can also be derived in the cases of very high and very low SNR.

When $\frac{E}{{N_0 }} \gg 1$, the argument of the $\cosh$ in (\ref{M1_metric}) is much greater than $1$ so that the function $\cosh(z)$ can be approximated as $\frac{1}{2}e^{\left| z \right|} $. Therefore, after dropping the constant $\frac{2 \sqrt{E}}{N_0}$, the high SNR approximation of the optimum path metric in (\ref{M1_metric}) can be expressed as
\begin{eqnarray}
\label{M2_metric}
    M_2 \left( {v_i^N } \right) &=& \sum\limits_{l = i}^N {\left(y_l s_l -\left| {y_l } \right|\right)}
\end{eqnarray}
which is much simpler than the metric $M_1$.

For practical implementation, after receiving $\mathbf{y}$, the values of $\left(y_l s_l -\left| {y_l } \right|\right)$  for all $s_l \in \{-1, +1\}$ can be pre-calculated and stored in a $2 \times N$-sized table. Thus, for each node visiting during the rest decoding procedure, no extra calculation is needed compared with the SSD using the conventional metric $M_0$. The similar method can also be applied for the calculation of metric $M_1$.

When $\frac{E}{{N_0 }} \ll 1$, the function $\log\cosh(z)$ can be approximated by $\frac{1}{2}z^2$. Using this approximation in (\ref{M1_metric}), we have
\begin{equation}\label{M3_metric}
M_3 \left( {v_i^N } \right) = \sum\limits_{l = i}^N {\frac{{2\sqrt E y_l s_l }}{{N_0 }}}  - \sum\limits_{l = i}^N {\frac{{2Ey_l^2 }}{{N_0^2 }}}  - \left( {N - i+1} \right)\log 2.
\end{equation}
But according to our simulation results (which are not provided in this paper), this metric is not efficient in reducing the complexity. So we will not discuss it below.

\subsection{The proposed SSD algorithm}
The SSD algorithm based on the proposed path metric $M_1$ or $M_2$ can be summarized as follows:

\medskip

\textbf{Step 1}. Initialization:
1) set the target radius as $r=\infty$;
%2) for all $i\in \{1,2,\cdots ,N\}$, in the case of $M_1$ metric, calculate $h_1\left(y_i^N \right)$, otherwise, in the case of $M_2$ metric, calculate $h_2\left(y_i^N \right)$;
2) a null path $\phi $ with $D\left( \phi  \right)=0$ is pushed into the stack.

\textbf{Step 2}. Popping: a path $v_i^N$ is popped from the top of the stack, if the path reaches the depth $N$, i.e., $i=1$, record the estimation vector $\hat{\mathbf{v}}=\mathbf{v}$ and update the search radius ${{r}}=\sqrt{D\left(v_i^N\right)}$, then go to \textbf{Step 4}.

\textbf{Step 3}. Expanding: the current path $v_i^N$ is extended to two new paths, i.e., $\left( 0,{{v}_{i}},{{v}_{i+1}},...,{{v}_{N}}, \right)$ and $\left( 1,{{v}_{i}},{{v}_{i+1}},...,{{v}_{N}} \right)$. For each path, calculate the SED $D\left(v_{i-1}^N\right))$ by (\ref{MSED_metric}). If the optimum path metric is used, calculate the path metric $M_1\left(v_{i-1}^N \right)$ by (\ref{M1_metric}). Otherwise, if the high SNR approximation is used, calculated the path metric $M_2\left(v_{i-1}^N \right)$ by (\ref{M2_metric}).

\textbf{Step 4}. Pushing: the paths with $D\left(v_{i-1}^N\right) < {{r}^{2}}$ are pushed back into the stack, the others are simply dropped.

\textbf{Step 5}. Sorting: the paths in the stack are re-sorted from top to bottom in descending order according to the value of $M_1$ or $M_2$.

\textbf{Step 6}. Stopping: if the stack is empty, then stop the decoding process and output $\hat{\mathbf{u}}=\hat{\mathbf{v}}\mathbf{B}$; otherwise, go to \mbox{\textbf{Step 2}}.

\medskip

To efficiently implement an SSD decoder, the same technique in \cite{scl} can be utilized to avoid the redundant copy operations and memory occupation. Compared with the conventional SD, the main difference of SSD algorithm is the sorting operations in \textbf{Step 5}. The candidate path is enumerated from the most probable one by applying the proposed path metric $M_1$ or its approximation $M_2$, therefore the complexity of SSD algorithm can be efficiently reduced.

\section{Simulation Results and Complexity Analysis}
\label{secIV}
In this section, the SSD is applied to polar codes and Reed-Muller (RM) codes with short code blocks.
The performance and complexity of the SSD with the conventional metric and the proposed metrics are evaluated via simulations with BPSK modulation over AWGN channels.

\begin{figure}[!t]
\centering{
\includegraphics[ width=0.9\columnwidth]{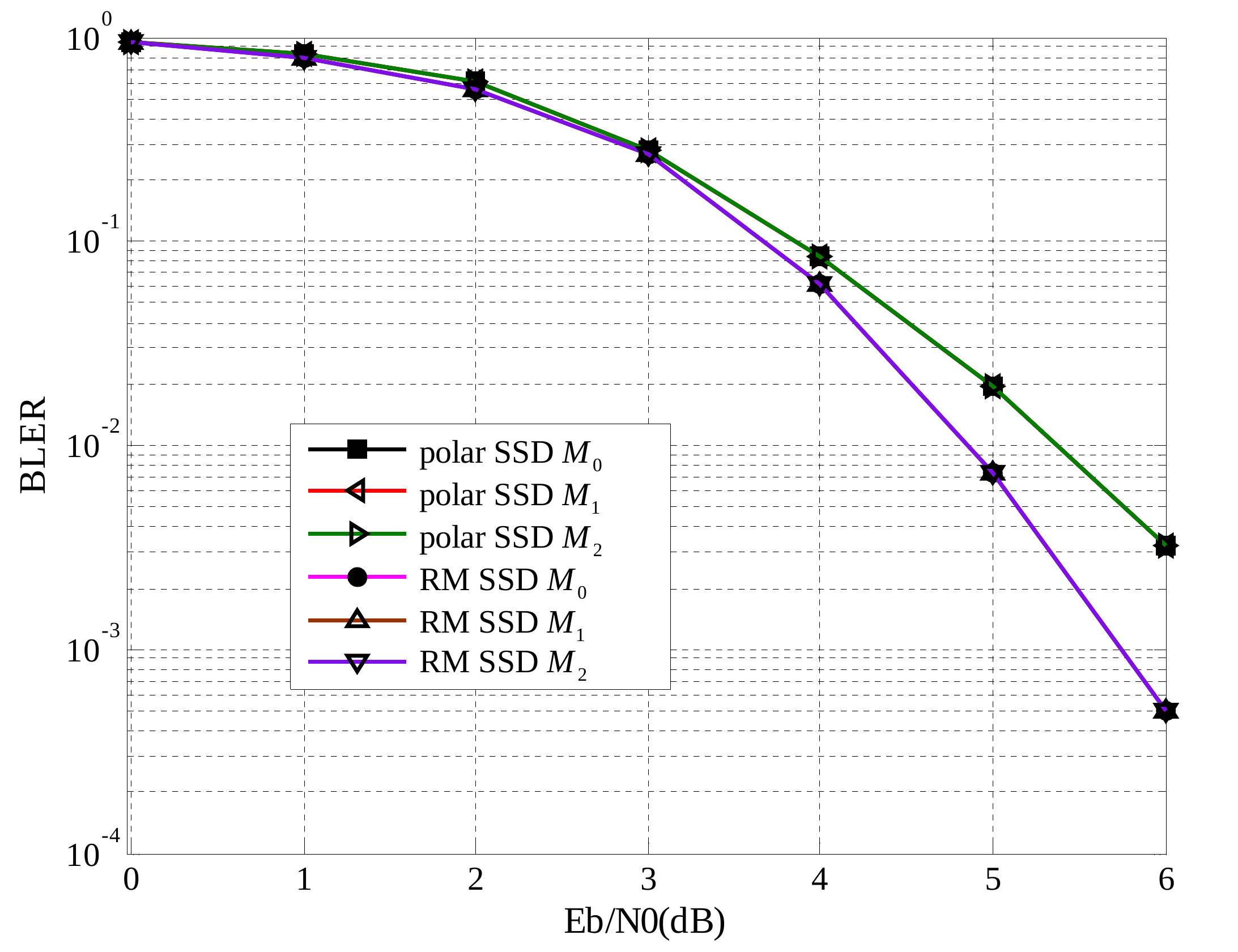}}
\caption{BLER performance of (64, 57) polar and RM codes over AWGN channels with SSD algorithms (some of the curves are overlapped).}
\label{fig_perform}
\end{figure}

Fig.\ref{fig_perform} presents the block-error-rate (BLER) performances of (64, 57) polar and RM codes under SSD with different path metrics.
We can see that the SSD algorithms with all the three metrics, the conventional metric ($M_0$ in (\ref{M0_metric})), the ML rule ($M_1$ in (\ref{M1_metric})) or its high SNR approximation ($M_2$ in (\ref{M2_metric})), can achieve the same performance.
Further, RM codes are shown to have better performances than polar codes under the ML decoding.

\begin{figure}[!t]
\centering{
\includegraphics[width=0.9\columnwidth]{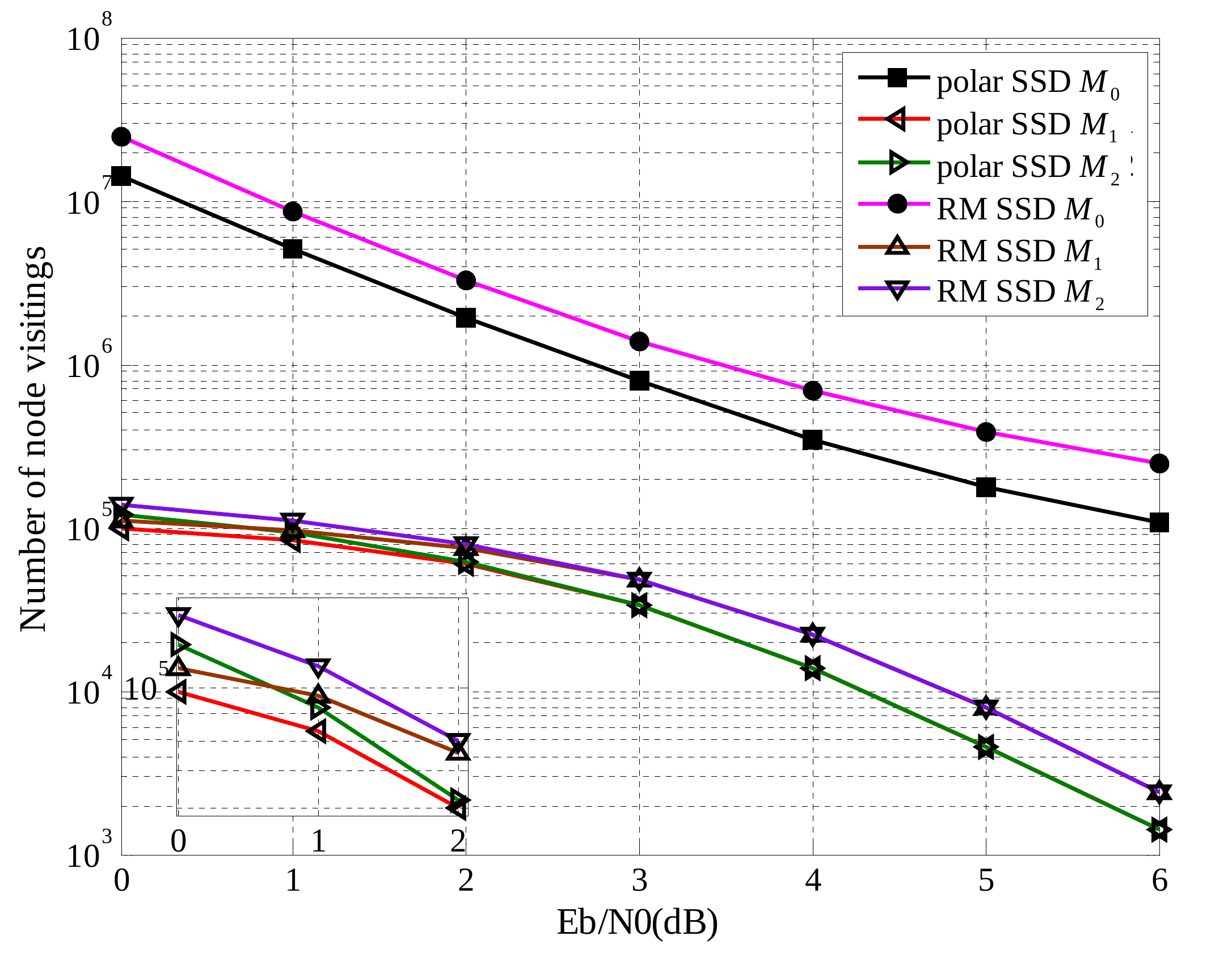}}
\caption{Average complexity of (64, 57) polar and RM codes over AWGN channels.}
\label{fig_complexity}
\end{figure}

The computational complexities are shown in Fig.\ref{fig_complexity}.
The average complexities are evaluated by counting the node visitings in the code tree.
Compared with the SSD using the conventional metric $M_0$, significant complexity reductions can be obtained by using the proposed path metrics: about $100$ times in the entire simulated SNR regime. Although the metric $M_2$ is an approximation of the metric $M_1$ in the case of high SNR, the SSD algorithm with this metric can achieve almost the same complexity reduction in the medium to high bit SNR regime, that is, ${{E}_{b}}/{{N}_{0}}=2\sim6\text{dB}$.
Interestingly, even in the low SNR regime, only a slight increasing of complexity under the SSD algorithm with the metric $M_2$ is observed.

\section{Conclusion}
\label{secV}
A stack-based sphere decoding algorithm with efficient enumeration is proposed to achieve the performance of ML decoding for the polar codes.
By introducing the optimum path metric in the sorting operations, remarkable complexity reduction can be obtained under the SSD algorithm.
Moreover, a high SNR approximation of the optimum metric is provided, which has a very simple form and is more suitable for practical applications.

\end{document}